\documentclass[prd,twocolumn,showpacs,preprintnumbers,amsmath,amssymb,floatfix]{revtex4}
\usepackage{graphics}
\usepackage{epsfig,float}
\usepackage{subfigure}

\begin{document}
\title{$\eta \to \pi^0 \gamma \gamma$ decay within a 
chiral unitary approach revisited}

\author{E. Oset$^1$, J. R. Pel\'aez$^2$ and L. Roca$^3$}
\affiliation{
$^1$Departamento de F\'{\i}sica Te\'orica and IFIC,
Centro Mixto Universidad de Valencia-CSIC, Institutos de
Investigaci\'on de Paterna, Aptdo. 22085, 46071 Valencia, Spain\\
$^2$Departamento de
F\'{\i}sica Te\'orica II,  Universidad Complutense. 28040 Madrid,
Spain.\\
$^3$Departamento de F\'\i sica, Universidad de Murcia,
E-30071, Murcia, Spain
}

\date{\today}

\begin{abstract}
In view of the recent experimental developments on the experimental
side in the $\eta \to \pi^0 \gamma \gamma$ decay, and the fact that
the Particle Data Group in the on line edition of 2007 reports
sizable changes of the radiative decay widths of vector mesons used
as input in the theoretical calculations of \cite{eta}, a
reevaluation of the decay width of the $\eta$ in this channel has
been done, reducing its uncertainty by almost a factor of two. The
new input of the PDG is used and invariant mass distributions and
total widths are compared with the most recent results from AGS,
MAMI and preliminary ones of KLOE. The agreement of the theory with
the  AGS and MAMI data is very good, both for the total rates as
well as for the invariant mass distributions of the two photons.
\end{abstract}


\maketitle

\section{Introduction}
The $\eta \to \pi^0 \gamma \gamma$ reaction has been quite controversial 
 given the large discrepancies between different theoretical approaches trying
 to match the scarce experimental data. 
For a long time the standard experimental results have been those of early 
experiments \cite{exp},\cite{Hagiwara:pw}, giving $\Gamma=0.84 \pm 0.18\,$eV. 
More recent experiments with  the Crystal Ball detector at AGS \cite{nefkens}
 reduced this value to
 $\Gamma=0.45 \pm 0.12\,$eV. A new reanalysis of AGS data gives 
 $\Gamma=0.285 \pm 0.031 \pm 0.049\,$eV \cite{prakhov}
  and a more recent analysis with the Crystal Ball at MAMI
  provides the rate
 $\Gamma=0.290 \pm 0.059 \pm 0.022\,$eV  \cite{prakhov}. At the same time
 the last two experiments  
 have provided the much awaited invariant mass distribution for the two photons,
 which was thought to provide valuable information concerning the theoretical
 interpretation. Some preliminary results from KLOE at Frascati \cite{Di Micco:2005rv}  
 are also
 available with values around $\Gamma=0.109 \pm 0.035 \pm 0.018$~eV.
 
 The theoretical models show also a similar dispersion of the results, 
 from large values obtained using models with quark box
diagrams \cite{Ng:sc,Nemoto:1996bh} to much smaller ones, obtained mostly using 
ideas of chiral perturbation theory (ChPT), which are quoted in 
 \cite{eta}.

The $\eta \to \pi^0 \gamma \gamma$ reaction has been traditionally considered to
be a border line
problem to test chiral perturbation theory (ChPT). The reason is that 
the tree level amplitudes, both at $O(p^2)$ and $O(p^4)$, vanish.
The first
non-vanishing contribution comes at $O(p^4)$, either from loops involving
kaons, 
largely suppressed due to the kaon masses, or from pion loops, again
suppressed since they violate G parity and are thus
 proportional to $m_u -m_d$ \cite{Ametller:1991dp}. 
The first sizable contribution comes at 
$O(p^6)$ but the coefficients involved are not precisely determined and one must
recur to models. In this sense, either Vector Meson Dominance (VMD) 
\cite{Ametller:1991dp,oneda,Picciotto:sn}, the 
Nambu-Jona-Lasinio model
(NJL) \cite{Bel'kov:1995fj}, or the extended Nambu-Jona-Lasinio model
(ENJL) \cite{Bellucci:1995ay,Bijnens:1995vg}, have
been used to determine these coefficients. However, the use of tree level VMD 
to obtain the $O(p^6)$ chiral coefficients
by expanding the vector meson propagators, leads \cite{Ametller:1991dp}
to results about a factor 
of two smaller than the "all order" VMD term when one keeps
the full vector meson propagator. The lesson obtained from these studies is that
ChPT can be used as a guiding principle but the strict chiral counting has to be 
abandoned since the $O(p^6)$ 
and higher orders involved in the full (``all order'') VMD results
are larger than those of $O(p^4)$. Also these calculations had several sources 
of uncertainty, one of the most important was the contribution of the $a_0(980)$
resonance, for which not even the sign was known. Thus, one is lead to rely
directly on mechanisms for the reaction, leaving apart the strict 
chiral counting.

 The theoretical situation improved significantly with the 
thorough revision  of the problem in  \cite{eta}, where the
different sources of uncertainty were studied and the $a_0(980)$
contribution was reliably included by using the unitary  extensions
of ChPT  \cite{npa,Kaiser:fi,ramonet,Nieves:2000bx}.  Within this
chiral unitary approach for the interaction of pseudoscalar mesons 
the $a_0(980)$, as well as the $f_0(980)$ or the $\sigma(600)$
resonances, are dynamically generated by using  as input the lowest
order chiral Lagrangians \cite{Weinberg:1966fm} and resuming the
multiple scattering series by means of the Bethe Salpeter
equation.  Another source of corrections in  \cite{eta} was the use
of the newest data for radiative decay of vector mesons of the PDG
2002 \cite{Hagiwara:pw}.  It was  noted in \cite{eta} that the
rates had significantly changed from previous editions of the PDG,
to the point that the $\eta \to \pi^0 \gamma \gamma$ widths
calculated in  \cite{Ametller:1991dp,Bijnens:1995vg} would have
changed by about a factor of two should one have used the new 
data for radiative decay of vector mesons of the PDG 2002 instead
of the former ones. Another improvement  in \cite{eta} was the
unitarization of the pair of mesons of the VMD terms beyond the
tree level. Furthermore, to have a better control on the reaction,
the consistency of the model with the related reaction $\gamma
\gamma \to \pi^0 \eta$ was established. Finally, in \cite{eta} a
thorough analysis of the theoretical errors was done by considering
all sources of uncertainty and making a MonteCarlo sample of
results obtained with random values of the input within the
uncertainties. 

   The final result obtained in \cite{eta} was 
\begin{equation}   
   \Gamma=0.42 \pm 0.14\,~eV,
   \label{eq:oldresult}
\end{equation}
which is still in agreement with
the present experimental results within uncertainties.
Nevertheless,  five years after the
publication of these results some novelties have
appeared that call for a revision of the problem. Indeed, once
again the data for the radiative decay of vector mesons of the "on
line" PDG 2007 \cite{pdg2007} have significantly changed with
respect to the data of the PDG 2002 used in ref.~\cite{eta}. The
correction due to these changes is important and it produces about
a 25\% decrease in the central value 
of the result of Eq.~(\ref{eq:oldresult}).
At the same time, the theoretical  uncertainty
is reduced by almost a factor of two.
On the other hand, the new experimental results regarding  the two
photon invariant mass distribution \cite{prakhov} provide an extra
challenge for the theoretical models.

In view of this, it has become necessary to update the work of 
\cite{eta} to account for the newest experimental results of the
PDG~2007 \cite{pdg2007} and to compare with the most recent
experimental data of the $\eta \to \pi^0 \gamma \gamma$ decay. The
model used here is, hence, the same as the one of \cite{eta} and
the only changes are the use as input of the new vector mesons
radiative widths. Thus, we refrain from providing detailed
explanations on the model and in this brief report we just
concentrate on the changes.

\section{VMD contribution}
\label{sec:VMD}

\begin{figure}[h]
\centerline{\hbox{\psfig{file=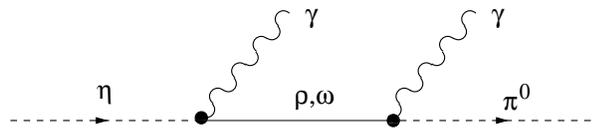,width=0.9\linewidth}}}
\caption{\rm 
Diagrams for the VMD mechanism.}
\label{fig:VMDtree}
\end{figure}

Following \cite{Ametller:1991dp} we consider the VMD
mechanism of Fig.~\ref{fig:VMDtree}
which can  be easily derived from the VMD Lagrangians involving
VVP and $V\gamma$ couplings \cite{Bramon:1992kr}
\begin{equation}
{\cal L}_{VVP} = \frac{G}{\sqrt{2}}\epsilon^{\mu \nu \alpha \beta}\langle
\partial_{\mu} V_{\nu} \partial_{\alpha} V_{\beta} P \rangle, \qquad
{\cal L}_{V \gamma} =-4f^{2}egA_{\mu}\langle QV^{\mu}\rangle,
\label{lagr}
\end{equation}
where $V_{\mu}$ and $P$ are standard $SU(3)$ matrices for the 
vector mesons  and  pseudoscalar mesons  respectively \cite{eta}.
In Eq.~(\ref{lagr}) $G=\frac{3g^2}{4\pi^2f}$,
$g=-\frac{G_VM_{\rho}}{\sqrt{2}f^2}$ \cite{Bramon:1992kr}
and $f=93\,$MeV,
 with $G_V$ the
coupling of $\rho$ to $\pi\pi$ in the normalization of
\cite{Ecker:1988te}.
From Eq.~(\ref{lagr}) one can obtain the radiative decay widths for
$V\to P\gamma$, which are given by
\begin{equation}
\Gamma_{V \to P\gamma}=\frac{3}{2} \alpha
 C_i^2 \left(G\frac{2}{3}\frac{G_V}{M_V}\right)^2k^3,
\end{equation}
where $k$ is the photon momentum for the vector meson at rest and
$C_i$ are $SU(3)$ coefficients that we give in Table~\ref{tab:BR} 
 for the different radiative decays, together
with the theoretical results
 (using $G_V=69\,$MeV and $f=93\,$MeV)
 and experimental \cite{Hagiwara:pw,pdg2007}
 branching ratios.
  In Table~\ref{tab:BR} we  quote the results of the PDG 
  version of 2002,
 which were used as input in the evaluation of the results
  in  \cite{eta}, together with the new results of the PDG~2007
  on-line edition \cite{pdg2007} which are used in the present 
  paper.

\begin{table*}[htbp]
\begin{center}
\begin{tabular}{|c||c||c|c|c|}\hline  
$i$ &$C_i$  &$B_i^{th}$ & $B_i^{exp}$ (PDG~2002 \cite{Hagiwara:pw})& $B_i^{exp}$ (PDG~2007 \cite{pdg2007})  \\ \hline \hline  
 $\rho\to \pi^0\gamma$  & $\sqrt{\frac{2}{3}}$  & $7.1\times 10^{-4}$ &$(7.9\pm 2.0)\times 10^{-4}$ &$(6.0\pm 0.8)\times 10^{-4}$ \\ \hline  
 $\rho\to \eta\gamma$ & $\frac{2}{\sqrt{3}}$  &  $5.7\times 10^{-4}$   &  $(3.8\pm 0.7)\times 10^{-4}$ &  $(2.7\pm 0.4)\times 10^{-4}$ \\ \hline  
 $\omega\to \pi^0\gamma$ &$\sqrt{2}$  & $12.0$\%  &  $8.7\pm 0.4$\% &  $8.91\pm 0.24$\% \\ \hline  
$\omega\to \eta\gamma$ & $\frac{2}{3\sqrt{3}}$ & $12.9\times 10^{-4}$  & $(6.5\pm 1.1)\times 10^{-4}$ &$(4.8\pm 0.4)\times 10^{-4}$ \\ \hline  
${ {K^{*+} \to K^+\gamma} \atop {K^{*-} \to K^-\gamma}}$ 
&$\frac{\sqrt{2}}{3}(2-\frac{M_{\omega}}{M_{\phi}})$  & $13.3\times 10^{-4}$  &  $(9.9\pm 0.9)\times 10^{-4}$ &  $(9.9\pm 0.9)\times 10^{-4}$ \\ \hline  
 ${ {K^{*0}\to K^{0}\gamma} \atop 
 {\overline K^{*0}\to \overline K^{0}\gamma} }$
 &$-\frac{\sqrt{2}}{3}(1+\frac{M_{\omega}}{M_{\phi}})$ & $27.3\times 10^{-4}$ & $(23\pm 2)\times 10^{-4}$& $(23.1\pm 2.0)\times 10^{-4}$ \\   
\hline
\end{tabular}
\caption{   SU(3) $C_i$ coefficients together with
 theoretical and experimental branching ratios for different 
vector meson decay processes. }
\label{tab:BR}
\end{center}
 \end{table*}

The agreement of the theoretical results with the data is fair but
they can be improved by incorporating $SU(3)$  breaking
mechanisms \cite{Bramon:1994pq}. For that purpose, we  normalize
here the $C_i$ couplings so that the  branching ratios in
Table~\ref{tab:BR} agree with experiment.

Once the $VP\gamma$ couplings have been fixed in this way,  we can
use them in the VMD amplitude corresponding to the diagram of
Fig.~\ref{fig:VMDtree}, for what we follow the details of \cite{eta}. Next we
briefly describe the other mechanisms considered in \cite{eta}.

\section{Other mechanisms}

 In \cite{eta} other mechanisms were considered which are not
affected by the modifications of the previous section. We refresh
them graphically and forward the reader to \cite{eta} for details.  

\begin{figure}[h]
\centerline{\hbox{\psfig{file=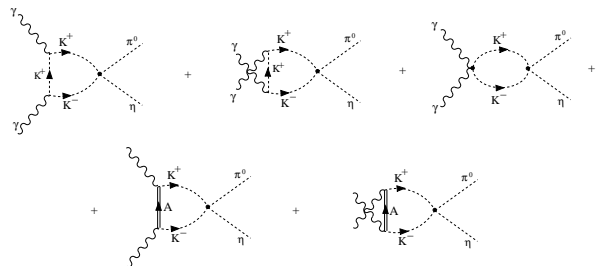,width=0.9\linewidth}}}
\caption{\rm
Diagrams for the chiral loop contribution  }
\label{fig.3}
\end{figure}

\begin{figure}[h]
\centerline{\hbox{\psfig{file=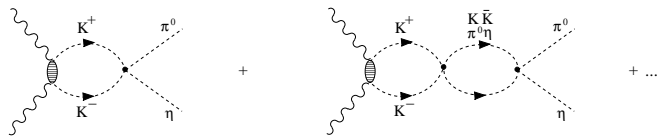,width=\linewidth}}}
\caption{\rm
Resummation for $\gamma\gamma\rightarrow\pi^0\eta$.}
\label{fig.5}
\end{figure}

\begin{figure}[h]
\centerline{\hbox{\psfig{file=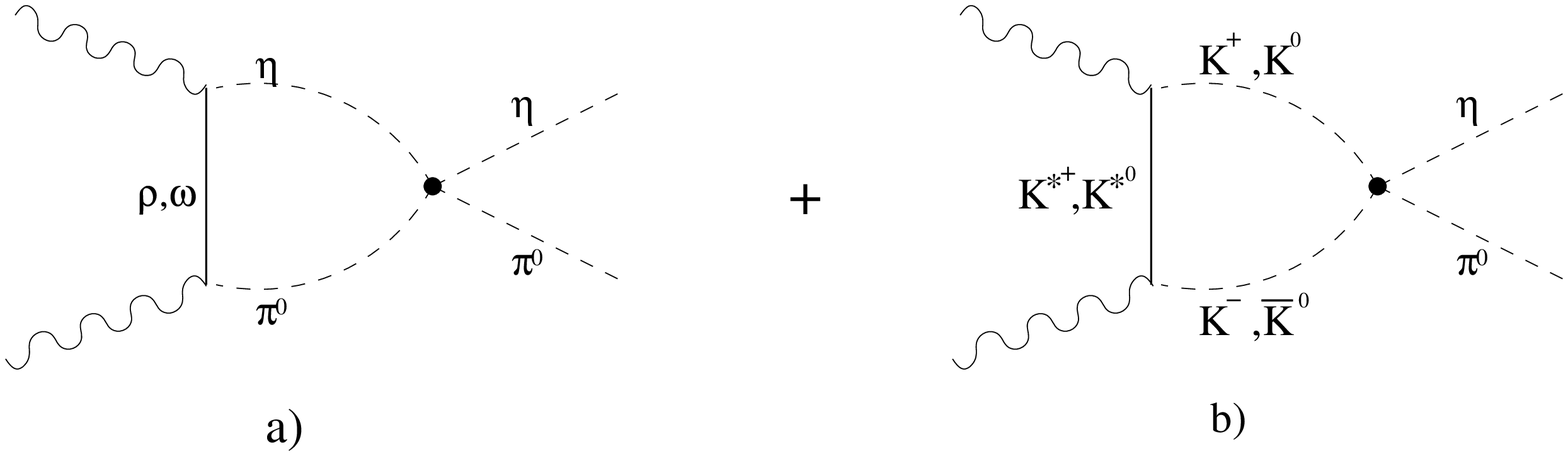,width=\linewidth}}}
\caption{\rm
Loop diagrams for VMD terms. The diagrams with the two crossed
photons are not depicted but are also included in the calculations.}
\label{fig.8}
\end{figure}

In Fig.~\ref{fig.3} we show the diagrams that go through kaon
loops. 
These diagrams, with the unitarization of the meson-meson
interaction depicted in Fig.~\ref{fig.5},
 were shown in
\cite{Oller:1997yg} to be mostly  responsible for the strength of
the
$\gamma \gamma \to \pi^0 \eta$ reaction in the region of the
$a_0(980)$ resonance. It was also shown in \cite{eta} that the
consideration of the mechanisms of Fig.~\ref{fig:VMDtree}
 and Fig.~\ref{fig.8} improved the agreement
with the data at low $\pi^0 \eta$ invariant masses.

 The vector meson exchange diagrams of Fig.~\ref{fig:VMDtree}
  were unitarized in \cite{eta}
 by including the resummation of diagrams of Fig.~\ref{fig.5},
 producing the diagram depicted in Fig.~\ref{fig.8}, where the thick
 dot represents the full meson-meson unitarized amplitude. Note that
 these mechanisms are also affected by the renormalization of the
 $VVP$ vertices discussed in the previous section.

\begin{figure}[h]
\centerline{\hbox{\psfig{file=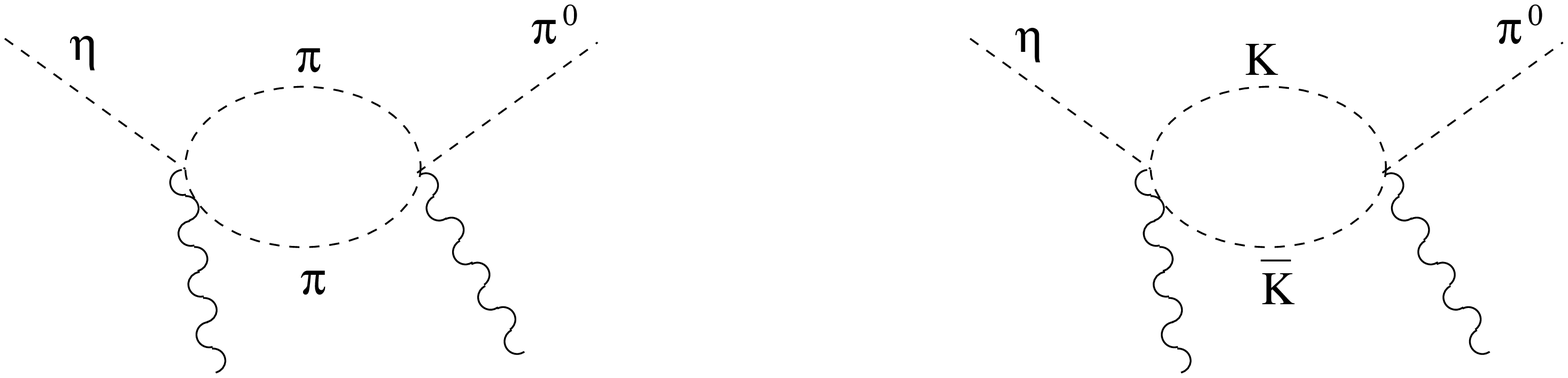,width=\linewidth}}}
\caption{\rm 
Diagrams with two anomalous $\gamma\rightarrow 3 M$ vertices.}
\label{fig.9}
\end{figure}

Finally, a small term related to the three meson axial anomaly,
 and shown
 diagrammatically in Fig.~\ref{fig.9},  was also taken 
 in the calculation since, as noted in \cite{Ametller:1991dp},
 although small by itself gives a non-negligible contribution upon
 interference with the other terms.
 
\section{Results}

By considering all the modifications discussed in
section~\ref{sec:VMD}, the integrated 
width that we obtain is 

\begin{equation}   
   \Gamma=0.33 \pm 0.08\,\textrm{ eV}
\end{equation}
which should be compared to the result of \cite{eta} of 
$\Gamma=0.42 \pm 0.14\,\textrm{ eV}$. The new result compares favorably with the 
most recent results of Cristal Ball at AGS  
$\Gamma=0.285 \pm 0.031 \pm 0.049\,\textrm{ eV}$ and MAMI 
$\Gamma=0.290 \pm 0.059 \pm 0.022\,\textrm{ eV}$ \cite{prakhov}. However, all these decay
widths are much larger than the preliminary results of KLOE at Frascati 
 $\Gamma=0.109 \pm 0.035 \pm 0.018$~eV \cite{Di Micco:2005rv}. 
 
\begin{figure}[h]
\centerline{\hbox{\psfig{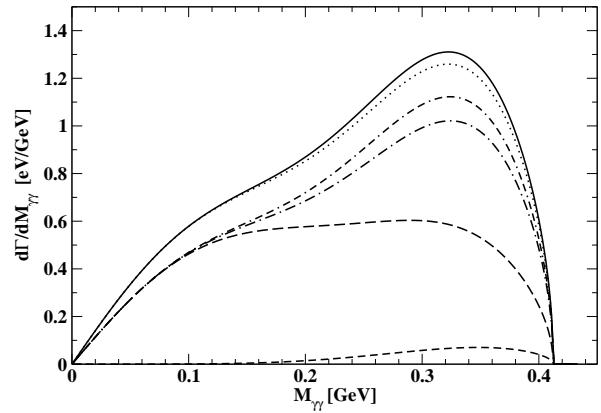}}}
\caption{\rm 
Contributions to the two photon invariant mass distribution.
 From bottom to top,
short dashed line: chiral loops; 
long dashed line: only tree level VMD;
dashed-dotted line:
coherent sum of the previous mechanisms; double
dashed-dotted line: idem  but adding the resummed VMD loops;
continuous line: idem but adding the anomalous terms of
Fig.~\ref{fig.9},
which is the full model presented in this work 
(we are also showing as a dotted line the
full model but substituting the full 
$t_{K^+K^-,\eta\pi^0}$
amplitude by its lowest order).
}
\label{mi}
\end{figure}

  The mass distribution of the two photons provides extra information which was
  claimed to be relevant to further test theoretical models. In \cite{eta} the
  differential cross section $d \Gamma / d M_{\gamma \gamma}$ was given. We
  present here the updated results in Fig.~\ref{mi}, where the contribution of
  the different mechanisms is shown. The new experiments reported in
  \cite{prakhov} provide measurements of $d \Gamma / d M^2_{\gamma \gamma}$
  which can be contrasted with theoretical predictions.
  
 Note that in the experiments of \cite{prakhov} 
  the magnitude $d \Gamma / d M^2_{\gamma \gamma}$
is given, while in \cite{eta} and in Fig. \ref{mi} 
$d \Gamma / d M_{\gamma \gamma}$ is evaluated. Although these distributions are 
equivalent, in practice the first one is more useful to study the spectrum at
low invariant masses since it provides extra information not given by the second
one. Indeed,  $d \Gamma / d M_{\gamma \gamma}$ is zero at the threshold of the
$\gamma \gamma$ phase space. However, 
$d \Gamma / d M^2_{\gamma \gamma}$ contains an extra 
$1/2 M_{\gamma \gamma}$ factor and leads to a finite value at zero
$\gamma \gamma$ invariant mass. This finite value and the shape
 of the
distribution close to threshold offer an extra test to the theory that would be
missed had we simply taken $d \Gamma / d M_{\gamma \gamma}$ for comparison. This
of course implies that the measurements can be done with good precision at the
threshold.  On the other hand, for the high mass region of the spectrum the 
$d \Gamma / d M_{\gamma \gamma}$ distribution is more suited to reveal the
effects of different theoretical mechanisms, as we have shown in Fig. \ref{mi}.

\begin{figure*}[h]
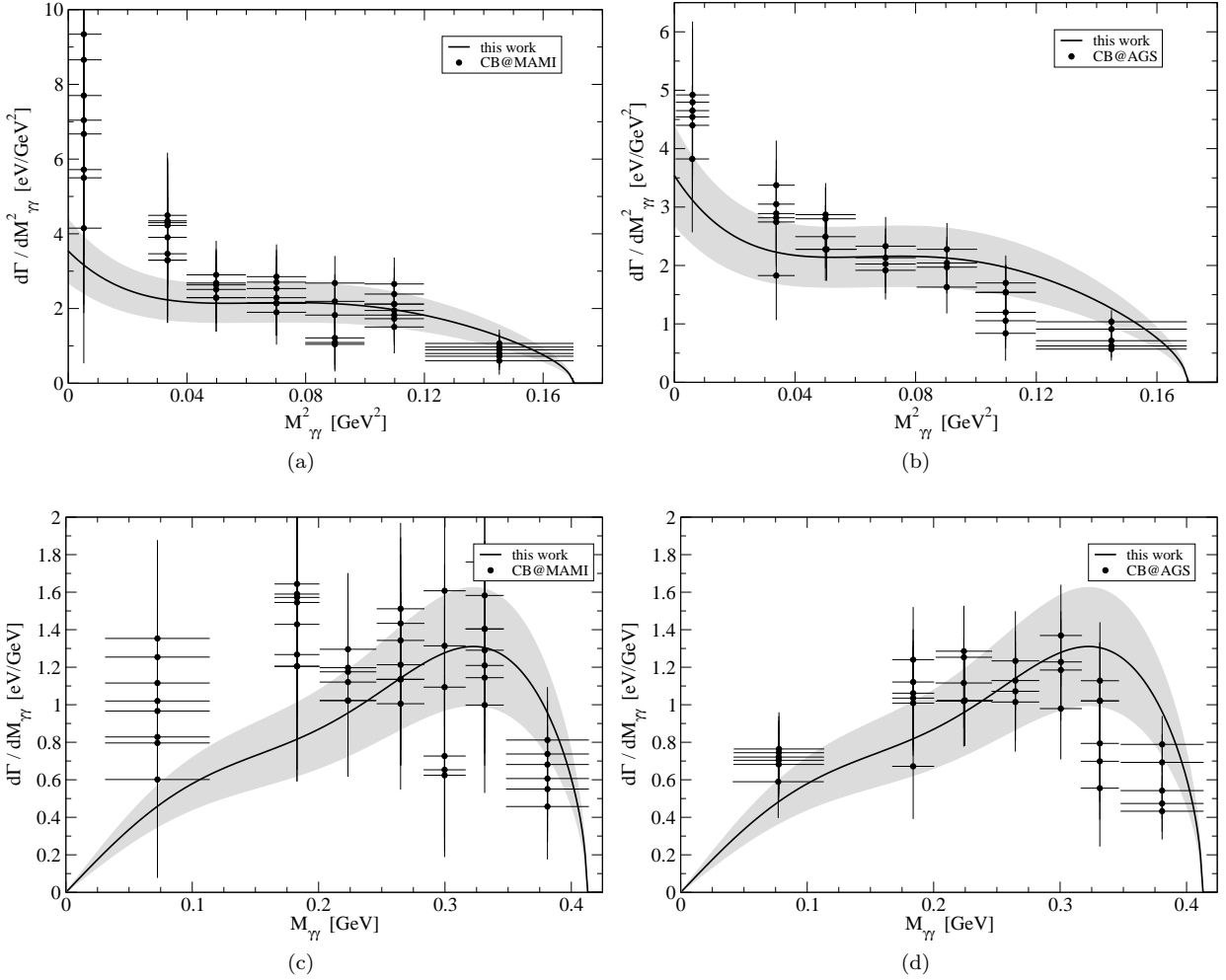

     \centering
     \subfigure[]{
          \label{mi2mami}
          \includegraphics[width=.45\linewidth]{figure7.eps}}
     \subfigure[]{
          \label{mi2ags}
          \includegraphics[width=.45\linewidth]{figure8.eps}}
   \\ \vspace{0.1cm}
       \subfigure[]{
          \label{mimami}
          \includegraphics[width=.45\linewidth]{figure9.eps}}
     \subfigure[]{
          \label{miags}
          \includegraphics[width=.45\linewidth]{figure10.eps}}
     \caption{Two photon invariant mass squared 
 (upper raw) and two photon invariant mass (lower raw) distributions.
 The data are from \cite{prakhov} for the Crystal Ball at MAMI
 (left panels) and for the Crystal Ball at AGS (right panels).
  The shaded region
 corresponds to the band of values of the present work considering the
 theoretical uncertainties.}
     \label{fig:mi}
\end{figure*}

In Figs.~\ref{mi2mami} and \ref{mi2ags} we compare the theoretical results that 
we obtain
with the distributions obtained at MAMI and AGS. 
  The agreement is good, in shape and size, and the theory provides indeed a
finite value at threshold compatible with experiment, which has nevertheless 
large errors. It is interesting to see that the AGS data show clearly an
increase of the distribution at low invariant masses which is a feature of the
theoretical results. The data of MAMI, however, have too large errors at
threshold and does not allow one to see this trend of the results. 
At large values of the invariant mass the agreement of the theory with the
MAMI data is better than with the AGS data.

  In order to offer a different perspective of the comparison of the results
at the higher mass region of the distribution,
 we show in Figs.~\ref{mimami} and \ref{miags} our final results for
  $d \Gamma / d M_{\gamma \gamma}$ compared to the data of \cite{prakhov}
  properly transformed to these variables.

\section{Summary}
In summary, we have witnessed an important experimental advance in the recent
years on the   $\eta \to \pi^0 \gamma \gamma$ decay. The parallel advances in
theory reflected by the work of \cite{eta} have allowed a detailed comparison of
results which has given a good agreement both for the total rate as well as for
the invariant mass distributions with the most recent finished results. The
discrepancy with the preliminary data of Frascati is worrisome, but we should
wait till these data are firm before elaborating  further  on the 
discrepancies.

\begin{acknowledgments}

This work is partly supported by DGICYT contract number
FIS2006-03438, and the Generalitat Valenciana. This research is 
part of the EU Integrated Infrastructure Initiative  HADRONPHYSICS
PROJECT Project under contract number RII3-CT-2004-506078. JRP's
research is partially funded by Spanish CICYT contracts
FPA2007-29115-E,FIS2006-03438, FPA2005-02327, UCM-CAM 910309, as
well as Banco Santander/Complutense contract PR27/05-13955-BSCH.
L.R. akcnowledges further support from  Fundaci\'on S\'eneca grant
No. 02975/PI/05 and CICYT contracts FPA2004-03470 and
FPA2007-62777.

\end{acknowledgments}

\end{document}